\documentclass{aa}

 \def        \pa         {\partial}
 \def        \grad       {{\bf \nabla}}
 \newcommand {\ddt}[1]   {{{\rm d} #1 \over{\rm d} t}}

\usepackage{graphics}

\begin{document}
 \thesaurus{ 09             
             (06.19.3;      
              06.13.1;      
              02.13.2)      
             \hfill\today 
           }

 \title{Magnetic flux tubes evolving in sunspots}
 \subtitle{A model for the penumbral fine structure and the
    Evershed flow}

 \author{R. Schlichenmaier\inst{1}\thanks{\emph{Present address:} 
            Kiepenheuer-Institut f\"ur 
            Sonnenphysik, Sch\"oneckstr. 6, D-79104 Freiburg; 
            schliche@kis.uni-freiburg.de}, 
            K. Jahn\inst{2}, H.U. Schmidt\inst{3}}
 \offprints{R. Schlichenmaier, Freiburg}

 \institute{Max-Planck-Institut f\"ur extraterrestrische Physik,
            D--85748 Garching, Germany;
       \and Warsaw University Observatory, Al. Ujazdowskie 4, 
            Pl--00-478 Warsaw, Poland; crj@astrouw.edu.pl
       \and Max-Planck-Institut f\"ur Astrophysik, Karl Schwarzschildstr. 1, 
            D--85748 Garching, Germany;}

 \date{Received 29 January 1998; Accepted 26 June 1998}

 \titlerunning{Magnetic flux tubes in sunspots}
 \authorrunning{Schlichenmaier, Jahn, Schmidt}

 \maketitle

 \begin{abstract}

 Assuming that the interchange convection of magnetic flux elements
 is the physical cause for the existence of filamentary penumbrae in
 sunspots, we investigate the behavior of an individual fibril embedded
 in the deep penumbra. The fibril is approximated by a thin magnetic flux
 tube which evolves dynamically in the environment given by the global
 magnetostatic model of a sunspot.

 Our simulation shows that the flux tube, initially positioned at the
 penumbra--quiet Sun boundary in the model, will rise through its deep
 penumbra developing a flow along the tube that points upward beneath
 the photosphere, and radially outward above the photosphere.  Our
 results suggest that a bright filament may be formed by an extended
 tail of a penumbral grain. Such filaments are optically thick, hotter
 than the surroundings, and elevated above a darker background.  An
 upflow in penumbral grains bends horizontally outwards above the
 photosphere and gradually cools down due to radiative losses leading to
 a tail that gradually darkens. The plasma flow inside the flux tube
 then becomes transparent and the tube constitutes a thin elevated flow
 channel, that can reproduce the observed features of the Evershed
 effect.  We present also a new acceleration mechanism for the Evershed
 flow.  It is demonstrated that a local surplus of gas pressure develops
 inside the tube as it rises through the specific (superadiabatic and
 magnetized) penumbral background.  The resulting gradient of the gas
 pressure can drive the flow along the tube.


 \end{abstract}


 \section{Introduction}

A sunspot penumbra seen in high resolution images reveals a small scale
fine structure (for a review, see Muller \cite{Mul92}). At a resolution
better than $0.5\arcsec$, penumbral grains and radially elongated bright
and dark filaments become visible. There is some observational evidence
that bright filaments consist of a few penumbral grains which are
radially aligned (Muller \cite{Mul73b}; Soltau \cite{Sol82}). Penumbral
grains exhibit proper motions and migrate inwards, i.e. toward the
umbra, with apparent velocities of 300--500\,m~s$^{-1}$ (Muller
\cite{Mul73a}). Muller describes penumbral grains as comet-like
structures, having a bright coma and a somewhat dimmer tail, which is
always directed radially outwards, i.e. away from sunspot center (see
also Fig.~2 in Tritschler et al. (\cite{Tri97})).  He states that
penumbral grains have a diameter of $0.35\arcsec$. However, this value
is close to the spatial resolution limit that was achieved in his
observations. Grossmann--Doerth \& Schmidt (\cite{Gro81}) find on the
basis of a statistical analysis of penumbral fine structure, that bright
and dark filaments have to be smaller than $0.55\arcsec$. Using the
1.5\,m MacMath telescope, Stachnik et al. (\cite{Sta83}) find spatial
scales of $0.11\arcsec$ within the penumbra.  These results rely on
power spectra which were obtained using speckle interferometry.  Thus,
since the spatial dimensions of bright and dark filaments are uncertain,
it is impossible to assign reliable values for corrected photospheric
intensities and effective temperatures of the penumbral fine structure.

A still not fully understood penumbral phenomenon was discovered by
Evershed (\cite{Eve09}): photospheric absorption lines show a shift of
the line core that turns out to be related to a line asymmetry when dark
penumbral structures are observed (Beckers \& Schr\"oter \cite{BuS69};
Wiehr \& Degenhardt \cite{WuD94}).  In recent years evidence has
accumulated that the Evershed effect must be due to a magnetized plasma
flow (Solanki et al. \cite{Sol94}) occurring in thin elevated horizontal
channels containing outward flows (Rimmele \cite{Rim95}).  The observed
asymmetry of the intensity profile can then be explained as the
superposition of an unresolved Doppler shifted component and an
unshifted main component. Here it is assumed that an individual flow
channel cannot be spatially resolved as suggested by many observers
(Bumba \cite{Bum60}; Holmes \cite{Hol63}; Schr\"oter \cite{Sro65};
Stellmacher \& Wiehr \cite{SuW80}; Wiehr \cite{Wie95}; Balthasar et
al. \cite{Bal97}).

The siphon flow mechanism has been proposed to explain the Evershed
effect by Meyer \& Schmidt (\cite{MuS68}). It was studied numerically
for stationary flux tubes, which arch from the inner penumbra to the
outer penumbral boundary, by numerous authors (e.g., Thomas
\cite{Tho88}; Degenhardt \cite{Deg91}; Thomas \& Montesinos
\cite{TuM93}). Hitherto, the stationary si\-phon flow model is the most 
broadly accepted explanation for the Evershed effect. For these models
it is essential to assume strong magnetic flux concentrations of
opposite polarity near the outer edge of the penumbra that can create a
gas pressure gradient along the arched flux tube in order to accelerate
the plasma. Here, we want to present for the first time a dynamical
model in which the plasma is accelerated locally in the penumbra. In our
model, the gas pressure gradient that accelerates the flow is naturally
caused by the evolution of the flux tube.

The subphotospheric structure of sunspot penumbrae cannot be observed
directly. It is deduced by means of numerical models for the overall
structure of a spot (see the review of Jahn \cite{Jah92}); models with
characteristics at the surface which are concurrent with observations
yield some information about the spot's properties in deeper layers
(e.g., Jahn \cite{Jah89}; Jahn \& Schmidt \cite{JuS94}).  Both models
and observations of the magnetic field strength and its inclination
within the penumbra (e.g., Beckers \& Schr\"oter
\cite{BuS69}; Schmidt et al. \cite{Smi92}; Title et al. \cite{Tit93};
Lites et al. \cite{Lit93}) indicate that the penumbra cannot be a
surface phenomenon.  Instead, it has to be deep (Schmidt \cite{Smi87},
\cite{Smi91}).

The averaged photospheric heat flux of the penumbra, $F_{\rm pu}\approx
0.75F_{\sun}$, is significantly higher than the corresponding umbral
heat flux, $F_{\rm u}\approx 0.23F_{\sun}$.  Jahn \& Schmidt
(\cite{JuS94}, hereafter JS) proposed a concept of interchange
convection of magnetic flux tubes to account for the surplus brightness
of the penumbra. Inspired by observations of bright and dark filaments
they conjectured that the magnetic field within the penumbra consists of
an ensemble of magnetic flux tubes. They surmised that inclined magnetic
flux tubes evolve dynamically within the penumbra and are able to
transport heat more efficiently than magnetoconvection in the umbra
which contains an almost vertical magnetic field. The onset of
interchange convection should depend on the slope of the magnetic field
lines: Imagine a perturbation acting on a bundle of magnetic field lines
being in magnetohydrostatic equilibrium. Since the component of the
buoyancy forces acting perpendicular to the tube increases as the flux
tube becomes more horizontal, there may exist an angle for which
buoyancy forces perpendicular to the tube resulting from a perturbation
overcome restoring magnetic forces. Then interchange convection of
magnetic flux tubes sets in (see also Rucklidge et al. \cite{Ruc95}).

 \begin{figure}
 \resizebox{\hsize}{!}{\rotatebox{-90}{\includegraphics{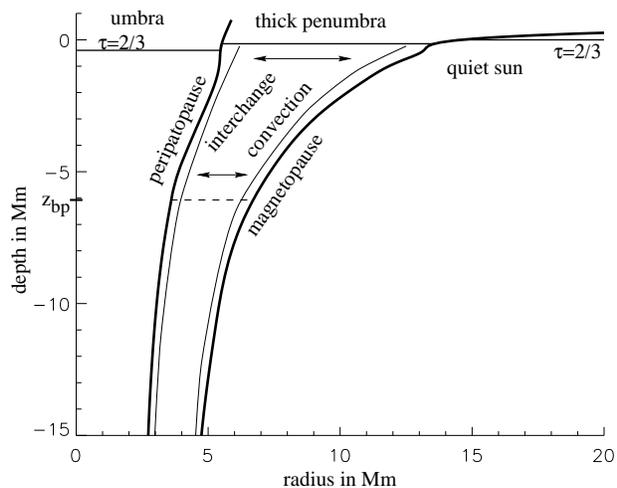}}}
 \caption[]{\label{figure1}
 {\bf Sunspot model in a meridional cut}. Above $z_{\rm bp}$, the
 magnetopause transmits heat which is distributed horizontally by the
 interchange of magnetic flux tubes. Both the peripatopause and the
 lower part of the magnetopause do not transmit energy (see also Fig. 2
 in JS).}
 \end{figure}

JS constructed the tripartite model for a sunspot which describes
self-consistently the three stratifications corresponding to the umbra,
the penumbra, and the quiet sun. These three stratifications are in
mechanical equilibrium, such that different gas pressures between the
adjacent stratifications are balanced by the magnetic pressure. The
corresponding magnetic fields are generated by two current sheets,
namely the magnetopause, which separates the quiet sun and the penumbra,
and the peripatopause, which separates the penumbra and the
umbra. Fig. \ref{figure1} shows a schematic drawing of these current
sheets.

All the stratifications are in hydrostatic equilibrium and the heat
transport (with varying efficiency) is described by means of the mixing
length theory. It is assumed that the peripatopause thermally insulates
the umbra, whereas heat exchange along the magnetopause is allowed in
subphotospheric layers as long as the slope of the magnetopause relative
to the vertical exceeds a critical angle of typically {30\degr}, which
corresponds to the bottom of the penumbra, $z_{\rm bp}$ (see
Fig. \ref{figure1}).  Since the energy transport is more efficient in
the quiet sun stratification than in the penumbra, the latter is heated.
The rate of heating depends on the transmissivity of the magnetopause,
which is a free parameter of the model. It is assumed that the
transmitted heat is distributed instantaneously on horizontal planes in
the penumbra model. Consequently, as it is adopted for the umbra and the
quiet sun, also the penumbral thermal stratification varies only with
depth.  Thus, the surplus brightness of the penumbra in the model is due
to a finite transmissivity of the magnetopause. The tripartite model of
JS has been successful in reproducing the observed magnetic flux of the
spot, the observed radial variation of the mean magnetic field strength,
the observed radii of sunspots, and the Wilson depressions of umbra and
penumbra. Moreover, interchange convection naturally causes an
inhomogeneous penumbra with features that are concurrent with many high
resolution observations (e.g., Schmidt et al. \cite{Smi92}; Title et
al. \cite{Tit93}; Lites et al. \cite{Lit93}).

The present investigation was started in order to test quantitatively
the concept of interchange convection.  Numerical investigation of the
dynamics of the whole ensemble of magnetic flux tubes is not possible at
present. As the natural first step, the discussion presented here is
devoted to the evolution of a single magnetic flux tube, which is
embedded in a tripartite sunspot.  Moreover, we assume the flux tube to
be {\em thin}.  The equations that govern the dynamics of a thin
magnetic flux tube are described in the next section.  We also describe
there briefly the numerical realization of the model. Sect. 3 describes
the model set-up, i.e. the initial conditions and the boundary
conditions, and the resulting evolution of a thin magnetic flux tube.
In Sect. 4, we compare the results of numerical simulations with the
observed photospheric features of penumbrae, and argue that our model
can reproduce the penumbral fine structure and the Evershed
effect. Also, the concept of interchange convection is discussed on the
basis of these new results. In the last section we summarize the results
emphasizing the observational signatures that are suggested by our
model.


 \section{The model}

We start by writing down the full 3-dimensional system of equations for
ideal magnetohydrodynamics (MHD) (see, e.g., Priest \cite{Pri82}). In
ideal MHD, the electrical conductivity is assumed to be infinite,
leading to frozen-in-magnetic fields. Using the substantial (convective)
derivative,
 
\begin{equation} \ddt{} = {\partial\over\partial t} +
\vec{v}\cdot \nabla \;\;, \label{substantial_derivative} 
\end{equation}
and cgs-units, the equations for continuity, induction, e\-quation of
motion, entropy equation, divergence free magnetic field, and the
equation of state read:

 \begin{eqnarray}
  \ddt{\rho} &=& -\rho \nabla\cdot {\bf v}
  \;\;,\label{continuity}\\
  {\partial\over\partial t}{\bf B} &=&\nabla\times ({\bf v}\times {\bf B}), 
  \;\;,\label{induction}\\
  \nabla\cdot {\bf B} &=& 0 
  \;\;,\label{divB}\\
  \rho \ddt{{\bf v}} 
  &=& -\grad p +{\bf g}\rho + {1\over{4\pi}} 
                    (\nabla\times{\bf B}) \times {\bf B}, 
  \;\;,\label{motion}\\
  \rho T \ddt{S} 
      &=& \rho T \left({{\rm d} S\over{\rm d} t}\right)_{\rm rad} 
  \;\;,\label{energy}\\
  p &=& {{\cal R}\over\mu} \rho T 
      + {4\sigma\over 3c} T^4
  \;\;.\label{state}
\end{eqnarray}
 
Here, $\rho$, $T$, $p$, $\vec{B}$, $\vec{v}$, and $\vec{g}$ have their
usual meaning. $S$ denotes entropy per unit mass, $\cal{R}$ the gas
constant, $\mu=\mu(\rho,T)$ the mean molecular weight. The second term
on the right hand side (RHS) in Eq.\ (\ref{state}) describes the
radiation pressure, with $\sigma$ being the Stefan--Boltzmann--constant,
and $c$ the speed of light. By taking advantage of the {\em thin flux
tube approximation} (Defouw \cite{Def76}; Spruit \cite{Spr81a},
\cite{Spr81b}; Ferriz Mas \& Sch\"ussler \cite{FuS89}) this
3-dimensional problem can be reduced to a 1-dimensional problem.  Doing
so, we closely follow the approach of Moreno Insertis (\cite{Mor86},
hereafter MI).  Furthermore, adopting the tripartite model of JS as a
background, we can describe the dynamics of a thin flux tube evolving in
a 2-dimensional plane ($x$,$z$).

 \subsection{Wal\'{e}n equation and equation of motion}

In a thin magnetic flux tube physical variables vary only along the
tube. Thus, the tube constitutes a curve in a plane, and it is
appropriate to use a Lagrange representation. As the independent
variable, we have chosen the integrated mass, $a$, along the tube.

The position of a mass element is specified by the vector:
$\vec{x}(a,t) = (x(a,t),z(a,t))$. Then the velocity and tangent
vectors are defined as follows:

 \begin{eqnarray}
  \vec{v}(a,t) &=& \left.{\partial\vec{x}(a,t)\over\partial t}\right|_a 
  \;\;,\label{velocity}\\
  \vec{l}(a,t) &=& \left.{\partial\vec{x}(a,t)\over\partial a}\right|_t
  \;\;.\label{tangentvector}
 \end{eqnarray}
Note, that $\left.\partial / \partial t (...)\right|_a$ is identical
with the substantial de\-riv\-a\-tive, ${\rm d}/{\rm d}t$, of Eq.\
(\ref{substantial_derivative}). The arc length, d$s$, of a mass element,
d$a$, is given by ${\rm d}s = l\cdot {\rm d} a$, with $l=|\vec{l}|$.

Within the thin flux tube approximation, the magnetic field vector in
the tube is always parallel to the tube,

 \begin{equation}
   \vec{B} = B \hat{\vec{t}}
 \;\;, \label{B_tangent}
 \end{equation}
with the tangent unit vector, $ \hat{\vec{t}} = \vec{l} / l $.
The magnetic flux, $\phi$, is conserved along the tube, i.e.,

 \begin{equation}
   \phi = BA = {\rm constant}
   \;\;, \label{flux_conservation}
 \end{equation}
 with $A$ being the cross section of the tube.  With the help of Eqs.
 (\ref{tangentvector}), (\ref{B_tangent}), and (\ref{flux_conservation})
 one immediately gets,

 \begin{equation}
    { \vec{B} \over \rho} = \phi  \vec{l}
    \;\;. \label{B_over_rho}
 \end{equation}
Finally, one assumes that the total pressure equilibrium of the
tube with its surroundings is achieved instantaneously:

 \begin{equation}
   p + {B^2\over 8\pi} = p_{\rm b} + {B_{\rm b}^2 \over 8\pi}
   \;\;.\label{pressure}
 \end{equation}
The index b denotes background variables. 

The continuity and the induction equations combined together yield the
Wal\'{e}n equation which can be written with a use of relations given
by Eqs. (\ref{tangentvector}) and (\ref{B_over_rho}) in the following
form:

 \begin{equation}
  \ddt{{\bf l}} = (\vec{l}\cdot\nabla)\vec{v} 
               = {\partial {\bf v} \over \partial a}  
  \;\;.\label{walen} 
 \end{equation}

 \begin{figure}
 \resizebox{\hsize}{!}{\includegraphics{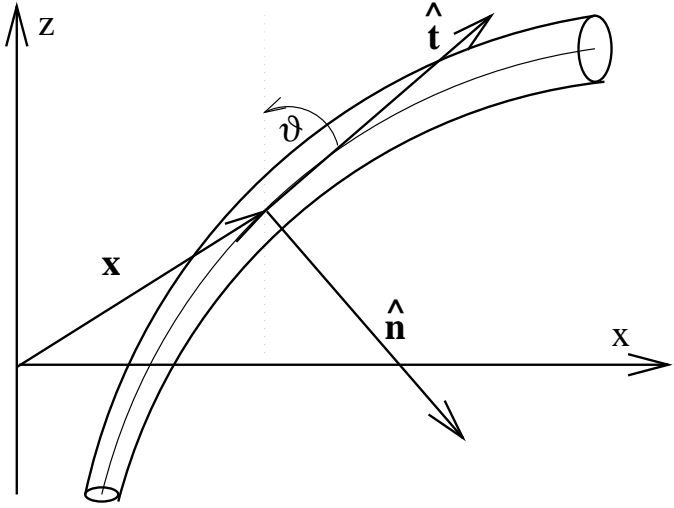}}
 \caption[]{\label{figure2}{\bf Geometric variables.}  The motion of a
 curved tube within a 2-dimensional plane can be characterized by the
 position vector {\bf x}$(a,t)$. The independent variable $a$ denotes
 the integrated mass along the tube. $\hat{\bf t}$ denotes the
 tangent unit vector, $\hat{\bf n}$ the normal unit vector, and
 $\vartheta$ the angle between the tangent vector and the vertical. }
 \end{figure}

The third term of the equation of motion (\ref{motion}) can be
simplified using the vector identities:

 \begin{eqnarray}
   (\nabla\times{\bf B})\times{\bf B}
  &=& B^2\kappa{\bf\hat n} 
     + {1\over 2}{\bf\hat t}{\partial\over\partial s}B^2
     - {1\over 2}\nabla B^2 
 \;\;,
 \end{eqnarray}
where $\hat{\vec{n}}$ denotes the normal unit vector as depicted in
Fig. \ref{figure2}, and the curvature $\kappa$ is defined as

 \begin{equation}
 \kappa {\bf\hat n} = {\partial{\bf\hat t}\over\partial s}
 \;\;.\label{kruemmung}
 \end{equation}

Then, using the condition of the instantaneous pressure equilibrium
(Eq. \ref{pressure}), the equation of motion can be rewritten:

 \begin{eqnarray}
   \lefteqn{ \ddt{{\bf v}}
    = {\bf g}\left(1-{\rho_{\rm b}\over\rho}\right) 
       +{{B^2\kappa}\over{4\pi\rho}} {\bf\hat n} 
       +{{\bf\hat t} \over 8\pi\rho}{\partial B^2\over\partial s}   
    }\nonumber\\
   && 
       -{1\over 8\pi\rho}\nabla B^2_{\rm b}
       +{F_D\over \rho} {\bf\hat n}	
   \label{pro_senk}
 \end{eqnarray}
The additional term on the RHS, $F_{\rm D}/\rho$, takes into account the
fluid flow around the tube. $F_{\rm D}$ denotes the drag force. We
follow the work of MI (see also  Caligari et al. \cite{Cal95}) and use

 \begin{equation}
   A F_{\rm D} = - \rho_{\rm b} {d \over 2} v_{\perp} |v_{\perp}|
   \;\;,
 \end{equation}
with $d$ denoting the diameter of the tube. 

It is appropriate to decompose the equation of motion into a component
parallel to the tube, and a component perpendicular to the tube. For the
parallel component, only the first two terms of Eq.\ (\ref{motion}) are
relevant. Hence,

 \begin{eqnarray}
  {\bf\hat t}\cdot \ddt{{\bf v}} = 
  \left({\rm d} {\bf v} \over {\rm d} t \right)_\| 
  &=& 
    - {1\over \rho l}{\partial p \over \partial a} 
    - g \cos\vartheta   
   \;\;,\label{dvpara}
 \end{eqnarray}
where $\vartheta$ denotes the angle between the vertical and the tangent
vector, $\hat{\bf t}$ (see Fig. \ref{figure2}). The perpendicular
component is obtained by multiplying the final form of Eq.\
(\ref{pro_senk}) with the normal unit vector $\hat{\bf n}$:

 \begin{eqnarray}
  \lefteqn{{\bf\hat n}\cdot \ddt{{\bf v}} = 
   \left({\rm d} {\bf v} \over {\rm d} t \right)_\perp = }
  \nonumber\\
  && 
   g\sin\vartheta\left( 1-{\rho_{\rm b}\over\rho}\right) 
   + {B^2\over 4\pi\rho}\kappa  
   - {{\bf\hat n}\over{8\pi\rho}} \cdot \nabla B^2_{\rm b} 
   + {F_{\rm D} \over \rho}
   \;\;. \label{dvsenk}
 \end{eqnarray}

Equations (\ref{walen}), (\ref{dvpara}), and (\ref{dvsenk}) determine
the time evolution of the tangent vector
$\vec{l}(a,t)=(l_x(a,t),l_z(a,t))$ and the velocity
$\vec{v}(a,t)=(v_x(a,t), v_z(a,t))$. Here, $v_x$ and $v_z$ are obtained
by projecting the velocity component according to
$\vartheta=\arctan{l_x/l_z}$.  These equations are identical to those
derived by MI, except that our tube is embedded in a magnetic
background.  Therefore, our perpendicular component of the equation of
motion contains an additional force (3$^{\rm rd}$ term on RHS of Eq.\
(\ref{dvsenk})), which results from the fact, that the gradient of the
magnetic pressure in the background exerts a force on the tube.  In the
magneto-hydrostatic background itself this force is canceled by the
magnetic curvature (2$^{\rm nd}$ term on RHS of Eq.\
(\ref{dvsenk})). Another subtle difference lies in Eq.\
(\ref{pressure}): in our case, the external pressure is given by the sum
of the background gas pressure and the magnetic pressure of the
background.

\subsection{Entropy equation}

The thermodynamic state of a mass element is characterized by $p(a,t)$,
$\rho(a,t)$, $T(a,t)$, and $B(a,t)$. These variables are determined by
three algebraic equations and one differential equation, i.e. Eqs.
(\ref{state}), (\ref{B_over_rho}), (\ref{pressure}), and (\ref{energy}).

The thermodynamic parameters of the flux tube are calculated in exactly
the same way as the thermodynamics in the background (see Jahn \cite{Jah89} for
details) including the effects of partial ionization on $\gamma =
c_p/c_V$ and on the mean molecular weight, $\mu=\mu(\rho,T)$ 
 (note that $\gamma$ and $\mu$ are not assumed to be constant).
The chemical composition is assumed according to Stix (\cite{Sti89}):
$X=0.7331$, $Z=0.01669$.

Using the first law of thermodynamics we can write the entropy change in
the following form:

 \begin{equation}
   {1\over c_V} \ddt{S} = \ddt{\ln T} - (\gamma-1)\cdot
      {\chi_\rho\over\chi_T}\cdot \ddt{\ln\rho} 
   \;\;.
 \end{equation}
 \begin{eqnarray}
 {\rm with }\qquad \chi_T &:=& \left.{\pa\ln p\over\pa\ln T}\right|_\rho
 \;\; , \label{chi_T} \\
 {\rm and }\qquad\; \chi_\rho &:=& \left.{\pa\ln p\over\pa\ln\rho}\right|_T   
 \;\;.  \label{chi_rho}
 \end{eqnarray}
This expression is substituted into the entropy equation
(\ref{energy}). After eliminating the total derivatives ${\rm d}\ln
p/{\rm d}t$ and ${\rm d}\ln T/{\rm d}t$, that result from
differentiating Eqs. (\ref{state}), (\ref{B_over_rho}), and
(\ref{pressure}), one finally obtains

 \begin{eqnarray}
   \lefteqn{\left(-{2\over\beta}-\gamma\chi_\rho\right) \ddt{\ln\rho} =}
   \nonumber \\
   &&{\chi_T \over c_V} \left(\ddt{S}\right)_{\rm rad}
   - {1\over p}\ddt{P_{\rm b}} + {2\over \beta} \ddt{\ln l}
   \;\; .\label{dlnrho}
 \end{eqnarray}
Note that the background is hydrostatic, $\partial P_{\rm b}/ \partial
t=0$, so that ${\rm d}P_{\rm b}/{\rm d} t = ({\bf v}\cdot{\bf \nabla})
P_{\rm b}$, i.e., the temporal change of $P_{\rm b}$ is only due to the
motion of a Lagrange point in the 2D plane.

 \subsubsection{Heat exchange by radiation}

Since radiation is the dominant physical process in the solar
photosphere, radiative heat exchange of the tube with the surroundings
has to be taken into account. We do this by adopting the relaxation time
approach by Spiegel (\cite{Spi57}, also used by Degenhardt \cite{Deg91}
and Thomas \& Montesinos (\cite{TuM93})). For a homogeneous atmosphere
in local thermodynamic equilibrium, the inverse damping factor of a
temperature perturbation can be identified with a radiative relaxation
time,
\begin{equation}
 t_{\rm rad} = \frac{c_p}{16\tilde\kappa \sigma T^3}\; [1-\tau\cdot {\rm
               arccot}(\tau)]^{(-1)} \;\; , \label{trad} 
\end{equation} 
where $\tilde \kappa$ is the Rosseland mean opacity (Huebner et
al. \cite{Hue77}). For small differences between the two temperatures,
the entropy change by radiation is then given by,
\begin{equation} 
 \left( \ddt{S} \right)_{\rm rad} =
 \frac{c_p}{t_{\rm rad} T} (T_{\rm b} - T) \;\;. \label{ds_rad}
\end{equation} 

 \subsection{Numerical treatment}

The evolution of a thin magnetic flux tube is given by 5 differential
equations (two components of Eq.~(\ref{walen}), (\ref{dvpara}),
(\ref{dvsenk}), and (\ref{dlnrho})) and 3 algebraic equations
((\ref{state}), (\ref{B_over_rho}), and (\ref{pressure})), which are
solved for the 8 unknown variables $\vec{l}(a,t)$, $\vec{v}(a,t)$,
$p(a,t)$, $\rho(a,t)$, $T(a,t)$, and $B(a,t)$. The tube's curvature,
$\kappa$, is given by

 \begin{equation} 
 \kappa = {1 \over l^3}\cdot \left( l_z {\pa l_x \over \pa a}  
        - l_x {\pa l_z \over \pa a} \right)
 \;\;.\label{kruemmung_num} 
 \end{equation} 

It is appropriate to use a staggered grid (see MI), in which $\vec{x}$
and $\vec{v}$ are discretized on the boundaries of a cell, and the
thermodynamical state, i.e. $p$, $T$, $\rho$, $B$, and $\bf l$ in the
center of the cells. The time integration is done explicitly in two
steps. First, the quantity $f$ is calculated for an intermediate step,
$f^{n+1/2}=f^{n}+(\ddt{f})^{n}\cdot{\rm d}t/2$. Then, the full time step
is calculated, $f^{n+1}=f^{n}+(\ddt{f})^{n+1/2}\cdot{\rm d}t$. The time
step used adapts itself, securing that the
Courant-Friedrichs-Lewy-criterion for stability is satisfied for each
cell, $dt \le \min({\rm d}s/ v_s, {\rm d}s/v_{\rm A})$. Here, ${\rm d}
s$ denotes the arclength of one cell, $v_s=\sqrt{\chi_\rho c_p/c_V}\cdot
\sqrt{p/\rho}$ the sound velocity, and $v_{\rm A}=B/\sqrt{4\pi\rho}$ the
Alfv\'en velocity. Simultaneously, the time step is forced to be smaller
than the radiation relaxation time, $t_{\rm rad}$. Once,
$\vec{l}^{n+1}$, $\vec{v}^{n+1}$, and $\rho^{n+1}$ are known, $B^{n+1}$
is determined by Eq. (\ref{B_over_rho}) and the position vector,
$\vec{x}^{n+1}$, that fixes the background variables, is obtained by
integrating the velocity in time. With $B^{n+1}$, $p_{\rm b}^{n+1}$, and
$B_{\rm b}^{n+1}$, Eq. (\ref{pressure}) yields the gas pressure,
$p^{n+1}$. The Saha equations determine the ionization fractions of
hydrogen and helium, and thereby the mean molecular weight $\mu$ and the
specific heat ratio $\gamma$. The proper pair of $\mu^{n+1}$ and
$T^{n+1}$ is found by iterating the equation of state, given by
Eq. (\ref{state}). The Saha equations and the iteration scheme are
described in more detail in Schlichenmaier (\cite{Sli97a}).

 \subsection{Embedding the tube in a tripartite model}

To follow the evolution of the flux tube in the penumbra we embed it in
a tripartite sunspot model. Here, the idea is that we take a bundle of
magnetic field lines from the sunspot model. These field lines are
initially in magneto-hydrostatic equilibrium and form a physical entity,
being described by a thin magnetic flux tube in ideal MHD. The
parameters of the particular tripartite model we used are listed in
Table \ref{parameters_bck}, for an illustration see Fig.\ 1 in JS.

 \begin{table}
 \caption[]{\label{parameters_bck}Parameters of background model (cf. JS)}
 \begin{tabular}{lcl}
 \hline\noalign{\smallskip}
 Parameter & Symbol & Value \\
 \noalign{\smallskip}\hline\noalign{\smallskip}
 total magnetic flux             & $\Phi_{\rm tot}$&$1\cdot 10^{22}$ Mx  \\
 magnetic flux of penumbra       & $\Phi_{\rm pu}$&$0.75\,\Phi_{\rm tot}$ \\
 Wilson depression of umbra 
 $^\dagger$                      & $W_{\rm u}$&$-470$ km \\
 Wilson depression of penumbra   & $W_{\rm pu}$&$-150$ km \\
 radius of penumbra 
 at $z\!=\!W_{\rm pu}$           & $R_{\rm pu}$&$13\,500$ km \\
 radius of umbra 
 at $z\!=\!W_{\rm u}$            & $R_{\rm u}$&$5000$ km \\
 transmissivity                  & $\epsilon$&$0.7$ \\
 lower boundary of model         & $z_{\rm bot}$&$-20\,000$ km \\
 upper boundary of model         & $z_{\rm top}$&$800$ km \\
 bottom of penumbra              & $z_{\rm bp}$&$-6\,000$ km \\
 $B$ at the  
 bottom of model                 & $B_{\rm bot}$&$15$ kG \\
 $B$ at the center in 
 photosphere                     & $B_{\rm cen}$&$2700$ G \\
 \noalign{\smallskip}\hline\noalign{\smallskip}
 \multicolumn{3}{l}{ $^\dagger$ 
 $z=0$ km corresponds to the photosphere ($\tau=2/3$) in}\\
 \multicolumn{3}{l}{ the quiet sun and decreases inwards.}
 \end{tabular}
 \end{table}

 \begin{figure}
 \resizebox{\hsize}{!}{\includegraphics{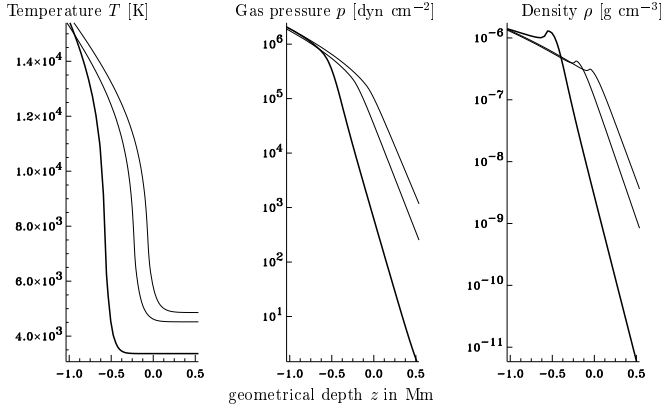}}
 \caption[]{\label{figure3}
 {\bf Thermodynamic variables of the background}. A sunspot is approximated
 by three different stratifications, i.e., the umbra (thick line), the
 penumbra (regular line), and the quiet sun (thin line). 
 }\end{figure}

The thermodynamic stratifications for the umbra, the penumbra, and the
quiet sun are prescribed by the tripartite model.  Within these
stratifications the pressure, $p_{\rm b}(z)$, the density, $\rho_{\rm
b}(z)$, and the temperature, $T_{\rm b}(z)$ depend only on depth.  Level
$z=0\,$km corresponds to $\tau=2/3$ level of the quiet sun and
decreases downwards.  Fig. \ref{figure3} shows the dependences of
$T_{\rm b}$, $p_{\rm b}$ and $\rho_{\rm b}$ on $z$ for the different
stratifications in the vicinity of the photosphere.
%
%
The thermodynamic variables inside the tube, at the center of each cell,
are obtained with a use of logarithmic splines defined on the background
discretization. The tube lying along the magnetopause is illuminated by
the quiet sun on one side. This effect is taken into account in the
radiative heat exchange.

 \begin{figure}
 \resizebox{\hsize}{!}{\includegraphics{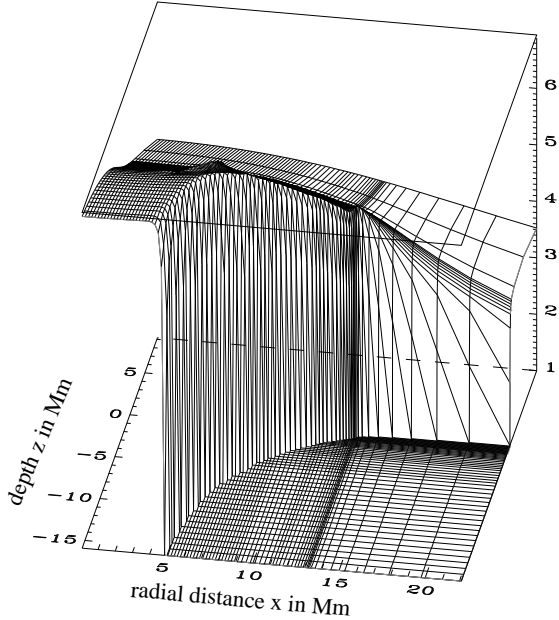}}
 \caption[]{ \label{figure4}
 {\bf The background magnetic pressure.} $B_{\rm b}^2/8\pi$, is shown as a
 logarithmic surface plot versus the radial distance, $x$, and depth, $z$.
 For the quiet sun, we set $\log(0)=1$.
 }\end{figure}

The background magnetic field strength, $B_{\rm b}(x,z)$, depends on the
radial distance, $x$, and on depth, $z$. The radial distance $x$ is
defined to be zero at the center of the sunspot model.  $B_{\rm b}(x,z)$
for a certain position is determined using a bilinear interpolation. The
magnetic pressure, $B^2_{\rm b}/8\pi$, of the used background model is
shown in Fig. \ref{figure4} as a surface plot. One can clearly see the
jump across the magnetopause from finite field strength in the penumbra
to a vanishing field in the quiet sun. The current sheets, i.e. the
magnetopause and the peripatopause, constitute discontinuities for the
magnetic pressure, $B^2_{\rm b}/8\pi$, the gas pressure, $p_{\rm b}$,
the temperature, $T_{\rm b}$, and the density, $\rho_{\rm b}$. But note,
that when crossing the current sheets horizontally, the total pressure
(gas plus magnetic) is constant.


 \section{The evolution of a thin magnetic flux tube embedded in a
          sunspot}

An ensemble of magnetic flux tubes is assumed to participate in
interchange convection. As a first step we restrict the study to
simulations of the evolution of a single flux tube. In this
contribution, we present the results obtained for one specific flux
tube and concentrate mainly on its photospheric behavior.

 \subsection{Initial configuration}

The results presented in this section have been obtained for the tube
with a magnetic flux $\phi = 2\cdot 10^{16}$ Mx. That corresponds to a
diameter of $d = 2\cdot \sqrt{\phi/B\pi}= 50$~km, for a magnetic field
strength equal to 1000 G. Initially, the tube lies along the
magnetopause. Hydrostatic
equilibrium along the tube is satisfied since the distribution of the
density and temperature is identical with the penumbral stratification:
 \begin{eqnarray}
 t=0: & \rho(x,z)  =& \rho_{\rm b,\rm penumbra}(z)    \nonumber \\ 
 &T(x,z)   =&    T_{\rm b,\rm penumbra}(z)    \nonumber \\ 
 &p(x,z)    =&    p_{\rm b,\rm penumbra}(z)    \nonumber 
 \end{eqnarray}
Magnetostatic equilibrium, i.e., the cancellation of the second and
third term on the RHS of Eq.\ (\ref{dvsenk}), is an intrinsic property
of the tripartite model, and can be realized numerically with an
accuracy which is sufficient, but limited by the interpolation
procedures. The magnetic field strength is given by the total pressure
equilibrium across the magnetopause,
 \[ t=0: \; B(x,z)    =  \sqrt{8\pi}\sqrt{
   p_{\rm b,\rm quiet\;sun}(z) - p_{\rm b,\rm penumbra}(z)} \;\;. \]

The model-tube extends down to a depth of $z= -15$ Mm at a radial
distance of $x=4.8$ Mm. The upper end is placed at a radial distance of
$x=24$ Mm and at a height of $z=0.3$ Mm above the photosphere
($\tau=2/3$) of the quiet sun. The intersection of the tube with the
photosphere coincides initially with the outer edge of the penumbra
model.

 \subsubsection{Boundary conditions}

The lower end of the tube, which is placed at the local position of the
magnetopause at $z=-15$ Mm deep in the convection zone, is held fixed,
i.e. the mass element at the lower boundary has zero velocity for all
times. At the upper boundary, we apply a free boundary condition: The
time derivative of the tangent vector, ${\rm d}{\bf l}/{\rm d} t$, is
extrapolated and the total force at the last grid point equals the total
force at the last but one grid point. In case of an outflow, mass
elements reaching $x=27$ Mm are cut off. That implies that the number of
grid elements gradually decreases, if an outflow is present.

 \subsubsection{Numerical tests}

The results presented in this section were obtained using 1033 grid
points, being distributed equidistantly in mass for $z>z_{\rm switch}=-1$
Mm and equidistantly in arclength below that depth.  It has been tested
that the evolution is the same for finer grids and does not depend on
the particular choice of the depth, $z_{\rm switch}$ (Schlichenmaier
\cite{Sli97a}).

When the radiative heat exchange is switched off, entropy becomes a
constant of motion. Since the entropy of the system is given by its
thermodynamic properties, the conservation of entropy per discretized
cell can be checked independently from the code (Schlichenmaier
\cite{Sli97a}). For this run, the code preserves entropy to a relative 
accuracy of less than $10^{-4}$.

 \subsection{Time series}

 \begin{figure}
  \begin{center}
   \resizebox{!}{22.79cm}{\includegraphics{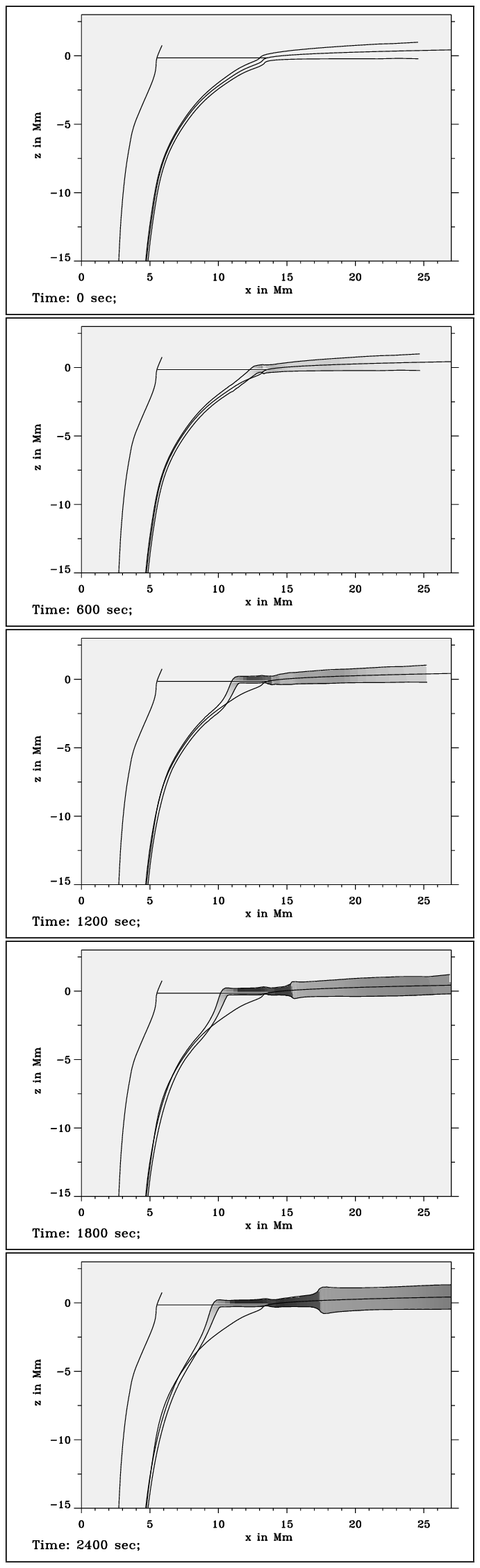}} \\
   \ref{time_series}a)
  \end{center}
 \end{figure}
 \begin{figure}
  \begin{center}
   \resizebox{!}{22.79cm}{\includegraphics{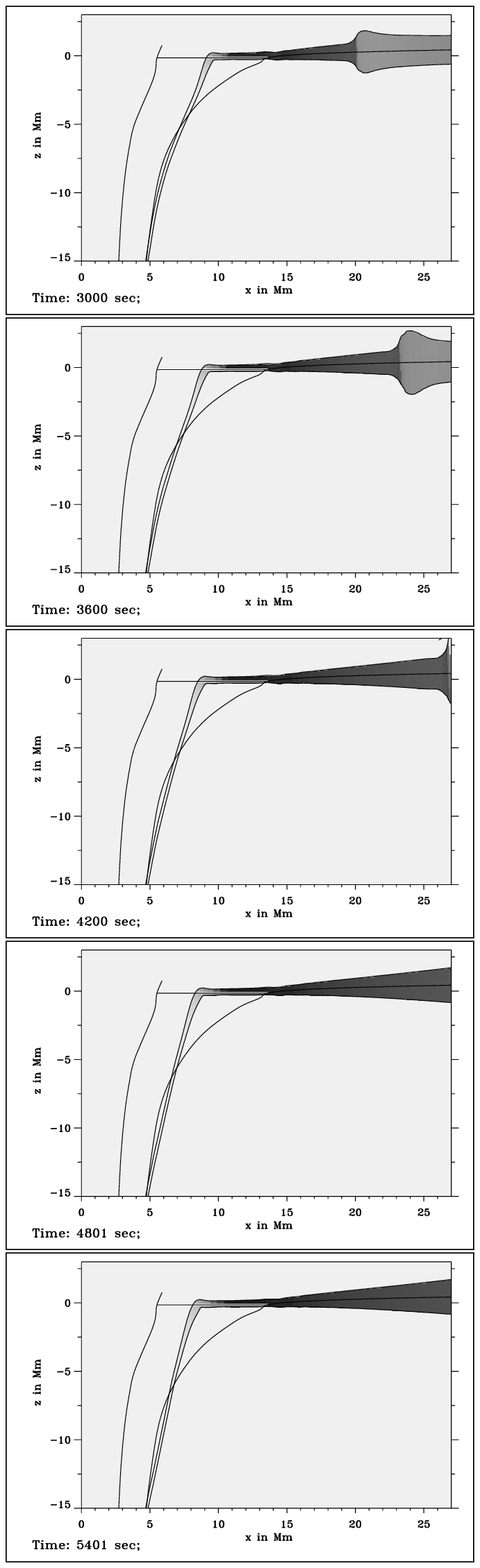}} \\
   \ref{time_series}b)
  \end{center}
 \end{figure}
 \begin{figure}
  \begin{center}
    \includegraphics{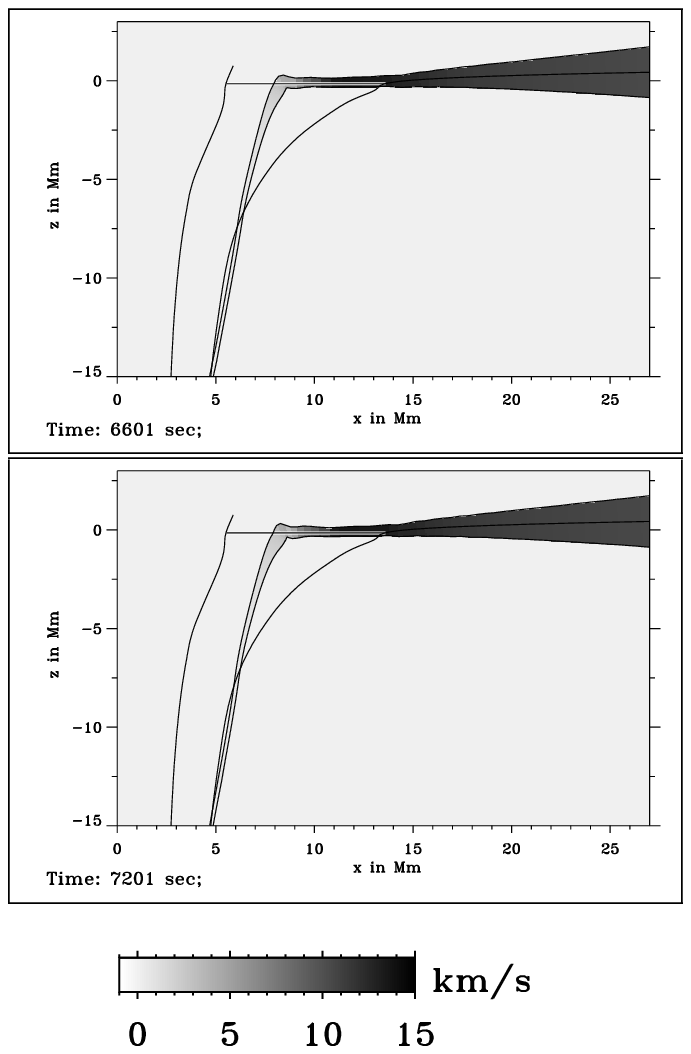}\\
    \ref{time_series}c)
  \end{center}
 \caption[]{\label{time_series}{\bf Time series} of the evolution of a
 flux tube that contains $\phi=2\cdot10^{16}$ Mx, and initially lies
 along the magnetopause. Starting just below the photosphere the tube
 rises, thereby causing the footpoint of the tube to migrate radially
 towards the umbra.  While it rises, a flow along the tube develops,
 being indicated by the gray-coding of the tube. The background-gray
 corresponds to 0 km~s$^{-1}$. The flow is directed upwards and
 outwards. At the outer edge of the penumbra the flow velocity becomes
 as high as 14 km~s$^{-1}$.}  
 \end{figure}

Fig. \ref{time_series} shows the time series of our simulation. Each
image shows the shape of the tube. The diameter of the tube has been
magnified by a factor of 12, for better visibility. The gray-coding
represents the velocity field along the tube. The background-shading
represents zero velocity, while darker coding corresponds to positive
velocities which point upwards and outwards. Images are plotted every
600 s.

 \subsubsection{The onset of evolution caused by radiation}

Even though initially the tube is in magnetohydrostatic equilibrium, it
will start to rise because of the radiative exchange of heat.  Part of
the tube is illuminated by the underlying hotter quiet sun so that it
heats up and expands relatively fast.  That leads to a decrease of
density inside the tube as compared to the outside, and the buoyancy
(first term on RHS of Eq.\ (\ref{dvsenk})) causes the tube to rise. The
region where the ascending motion appears first is determined by an
interplay between the radiative relaxation time, $t_{\rm rad}$, the
temperature difference, $T(z)-T_{\rm b}(z)$, and the superadiabaticity,
$\delta=\nabla -\nabla_{\rm ad}$. Since radiation is most effective near
the photosphere, the radiative relaxation time (see Eq. (\ref{trad})) is
the shortest there. Superadiabaticity, being the cause of convective
instability, is largest in the surface layers -- within the first few
hundred kilometers below the photosphere. The temperature difference
between the quiet sun and the umbra, being a measure for the rate of
heating, is largest near the photospheric level of the penumbra, i.e.,
at $z=-150$ km. Therefore, the tube gains buoyancy at the fastest rate
just below the photosphere and there the tube will start to rise first.
That can be seen in the second image of Fig.\ \ref{time_series}a.

 \subsubsection{Initial stages of the evolution}

Below the photosphere, the stratification is convectively unstable
(superadiabatic) and the tube's rise accelerates. Above the photosphere,
the stratification is convectively stable, and moreover, the tube's
density increases due to radiative losses. Therefore, the tube's rise
decelerates rapidly in the atmosphere. Approximately 100 km above the
photosphere, a new equilibrium is found for the tube, in which the
downward-acting buoyancy force is neutralized by the upward-acting
gradient of the background magnetic pressure (third term of RHS in Eq.\
(\ref{dvsenk})). Simultaneously, the footpoint of the tube (i.e. the
intersection of the tube with the photosphere at $\tau=2/3$), moves
radially inwards toward the umbra.

 \subsubsection{Build-up of surplus gas pressure inside the rising tube}
 \label{sec_surplus}

A positive flow velocity along the tube develops as the tube rises. It
reaches a maximum of 14 km~s$^{-1}$ at the outer edge of the penumbra at
later stages of the evolution. The onset of this longitudinal flow
results from an enhanced gas pressure gradient along the tube, which
builds up locally at the place where the tube rises into the
photosphere. In order to understand this phenomenon, one has to realize
that the tube finds itself (locally) in a strongly superadiabatic
environment characterized by a very small gas pressure scale height (of
the order of 100 km) which is much smaller than the scale height of the
magnetic pressure (approximately 3\,000 km). Note also, that below the
photosphere the gas pressure dominates the magnetic pressure,
i.e. $\beta>1$.

The tube stays in total (gas + magnetic) pressure equilibrium as it
rises (cf. Eq. (\ref{pressure})) and it must expand, because of the
decreasing pressure in the background. Since the decrease of the total
background pressure is dominated by the gas pressure, the magnetic
pressure, $B^2 / 8\pi$, inside the tube gets smaller than the background
magnetic pressure, $B_{\rm b}^2 / 8\pi$. In consequence, the gas
pressure inside, $p$, must build up relative to the background gas
pressure, $p_{\rm b}$. Note, that the background gas pressure is
constant horizontally within the penumbra.

Since the part of the tube lying above the photosphere does not rise, it
becomes almost horizontal between the footpoint, at $x_{\rm fp}$, and
the outer edge of the penumbra, at $x_{\rm oe}$, i.e. $p_{\rm b}(x_{\rm
fp}) \simeq p_{\rm b}(x_{\rm oe})$. Near the outer edge of the penumbra the
gas pressure inside the tube does not change, $p(x_{\rm oe}) = p_{\rm
b}(x_{\rm oe})$, whereas the gas pressure is greatly enhanced in the
vicinity of the footpoint, $p(x_{\rm fp}) > p_{\rm b}(x_{\rm fp})$,
since there the tube has risen through the strongly superadiabatic
region. The resulting gas pressure gradient accelerates a horizontal
outflow along the tube which is accompanied by an upflow in
subphotospheric layers. This upflow is maintained by the superadiabatic
background stratification. The gas pressure gradient in the horizontal
part of the tube is sustained by radiative losses which decrease the
internal energy (and the gas pressure) of the matter flowing along the
tube.

 \subsubsection{Centrifugal force and magnetic tension}

The migration of the footpoint is governed by three forces acting
perpendicular to the tube (the gradient of the background magnetic field
can be neglected here). First, at the turning point, i.e. slightly
above the footpoint, where the tube bends horizontally, magnetic
curvature force is positive (cf. Fig. \ref{figure2}) and it decelerates
the migration of the footpoint.  Second, the longitudinal flow
exerts a centrifugal force at the turning point. The centrifugal force
term stems from the advection term in
Eq. (\ref{substantial_derivative}), 
$\vec{v}\cdot \nabla$:

\[ \vec{v}\cdot \nabla\vec{v} = 
     \frac{\hat{\vec{t}}}{2} \frac{\partial v^2}{\partial s} 
    + v^2 \kappa \hat{\vec{n}} \;\;.\]
The first term gives the inertia force along the tube, and the second
term is the centrifugal force, which is the inertia force perpendicular
to the tube.

Third, the buoyancy being a vertical force that acts perpendicular to
the tube dominates the rise as long as the tube is highly inclined with
respect to the vertical.  As the time series shows, the inward
migration of the footpoint ceases at $t=3\,600$~s.  Here, the
buoyancy becomes negligible, because of the almost vertical orientation
of the subphotospheric part of the tube. Thus, in later stages of the
evolution, the perpendicular component of the equation of motion near
the turning point is governed only by the centrifugal and the magnetic
curvature forces. It is interesting to note, that the centrifugal
force neutralizes the magnetic curvature force when the flow velocity
equals the Alfv\'en velocity.

 \subsubsection{Formation of a shock front above the photosphere}

The time series shows a shock front that develops beyond the outer edge
of the penumbra. It forms at $x=15$ Mm after 1\,800~s and migrates
outwards with a velocity of roughly 3 km~s$^{-1}$. While it migrates
outwards and upwards, the shock front accelerates and reaches
$\approx$\,6 km~s$^{-1}$ at the outer (upper) boundary. Based on simple
physical arguments, the mere existence of this shock wave can be
explained as follows. Assume you have a stationary flow upwards along
the tube. Due to continuity, the flow velocity, $v_\|$, would then
increase according to $v_\| \sim 1/\rho$, whereas the Alfv\'en velocity
$v_{\rm A}$ scales as $v_{\rm A} \sim B/\sqrt{\rho}$. Now, $\rho$
decreases exponentially and, not so important, B decreases linearly with
height. Thus, $v_\|$ being smaller than $v_{\rm A}$ at some initial
height, increases more rapidly than $v_{\rm A}$. At some point, $v_\|$
has to become superalfv\'enic. In a magnetic flux tube the critical flow
velocity along the tube is given by the tube speed, $v_{\rm t}= v_{\rm
A}\cdot v_{\rm s}/ \sqrt{v_{\rm A}+ v_{\rm s}}$ (e.g., Ferriz Mas
\cite{Fer88}), implying that a superalfv\'enic flow is supercritical at
the same time.  Having a supercritical flow that collides with plasma
which is at rest initially has to result in a shock front. As this
shock-front accelerates the plasma, it migrates outwards and leaves the
computational domain after 4\,200~s.

In the present contribution we do not make any attempt to
realistically describe shock waves.  But it is clear that the shock
does not influence the dynamics happening near the penumbral
photosphere for several reasons. First, the shock front is a
{\it consequence} of the acceleration of plasma within the penumbra,
rather than the cause of the longitudinal flow. Second, the
density decreases exponentially with height, and therefore the inertia
within the photospheric penumbra is much to large to be affected by the
dynamics happening hundreds of kilometers above the photosphere. And
after all, the flow velocity upstream of the shock is supercritical,
i.e., the penumbral photosphere doesn't know about the shock front.

The flow is decelerated within the shock and the gas pressure increases
there.  That leads to a local expansion of the tube at the shock
front. Such an expansion should be counteracted by magnetic tension
forces inside the tube. However, within the thin flux tube approximation
such forces are neglected and nothing prevents the tube from infinite
expansion at the shock front. Therefore, we have included the magnetic
tension forces inside the tube fictitiously by changing the background
pressure, $P_{\rm b}$: At the location of expansion we enhance the
background pressure. The stronger the expansion, the larger the
background pressure.

 \subsubsection{Final stages of the simulated evolution}\label{finalkink}

 \begin{figure}
   \begin{center}
    \resizebox{\hsize}{!}{\includegraphics{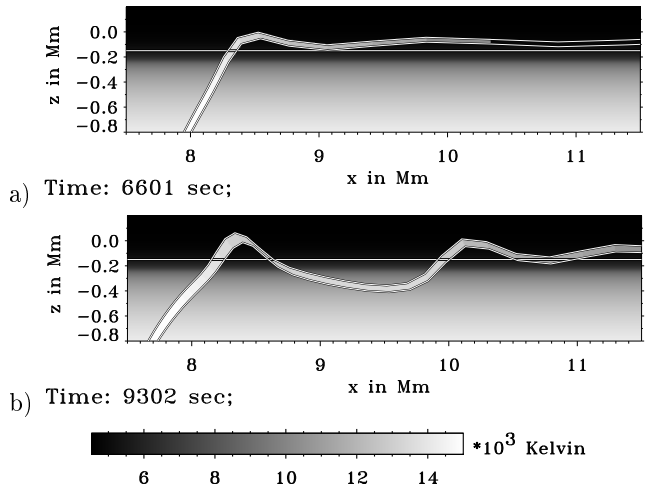}}
  \end{center}
 \caption[]{\label{abbcrash} {\bf Two snapshots of the flux tube and the
 temperature profile.}
 As the Alfv\'en velocity increases at the footpoint due to the continuous
 upflow along the tube, the centrifugal force at the turning point
 ($x\approx 8.4$ Mm) exceeds the magnetic tension, leading to overshooting.
 a) shows the phase in which the overshoot is convectively stable at $t=
 7\,500$ s, while b) visualizes the instable phase at $t=9\,300$ s, where
 the tube dives back into the subphotospheric stratification.  The gray
 coding refers to temperature according to the shading bar.
 }\end{figure}

As discussed in Sect. \ref{sec_surplus}, plasma within the tube
expands as it rises through the convection zone. Thereby, the magnetic
field strength, $B$, decreases. The magnetic field strength of plasma
reaching the footpoint is the smaller, the larger is the height
difference it traveled through the convection zone. That implies, that
$v_{\rm A}=B/\sqrt{4\pi\rho}$ at the footpoint decreases continuously,
while the flow velocity stays constant in time at the footpoint. Hence,
at some point (in our simulation at $t\approx 7\,000$~s) the flow
becomes superalfv\'enic and the centrifugal force exceeds the magnetic
tension force at the turning point, i.e. the flow overshoots the turning
point. Due to radiative losses in the atmosphere, the plasma that
overshoots gets denser and decelerates.

The amplitude of the overshoot gradually increases in time as
$v_\|/v_{\rm A}$ increases. Once the flux tube dives back into the
convectively unstable subphotospheric penumbra, anti-buoyancy drags
the tube down, and magnetic tension is too weak to prevent the tube from
sinking. Fig. \ref{abbcrash}b shows a snapshot of the tube in the
vicinity of the turning point after 9\,300 s. The inertia of the upflow
at the footpoint prevents the turning point from sinking. The part of
the tube that dives back beneath the photosphere sinks down. Since the
radii of curvature at the turning point ($x\approx 8.4$ Mm) and at
$x\approx 10.2$ Mm become very small, the thin flux tube approximation
is no longer valid. For that reason, we stop the simulation here.


 \section{Observational consequences}

 \begin{figure}
  \begin{center}
    \includegraphics{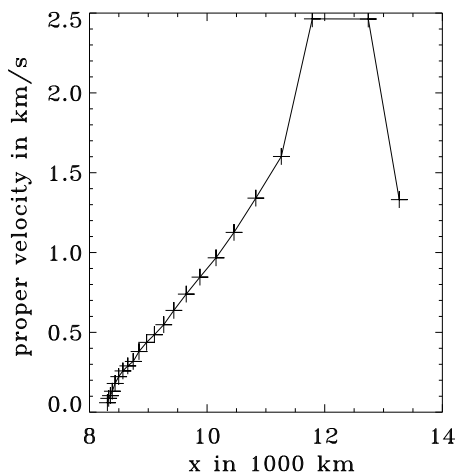}
  \end{center}
 \caption[]{\label{footpoint} {\bf Proper motion of the footpoint.}
 The apparent inward velocity of the migrating footpoint is plotted versus the
 radial distance $x$ from sunspot center. The plus signs are equally spaced
 in time ($\triangle t=300$s), tracing the evolution between $t=300$ s
 ($x=13\,300$ km) and $t=7\,200$ s ($x=8\,300$ km).}
 \end{figure}

\begin{figure*}
 \resizebox{\hsize}{!}{\includegraphics{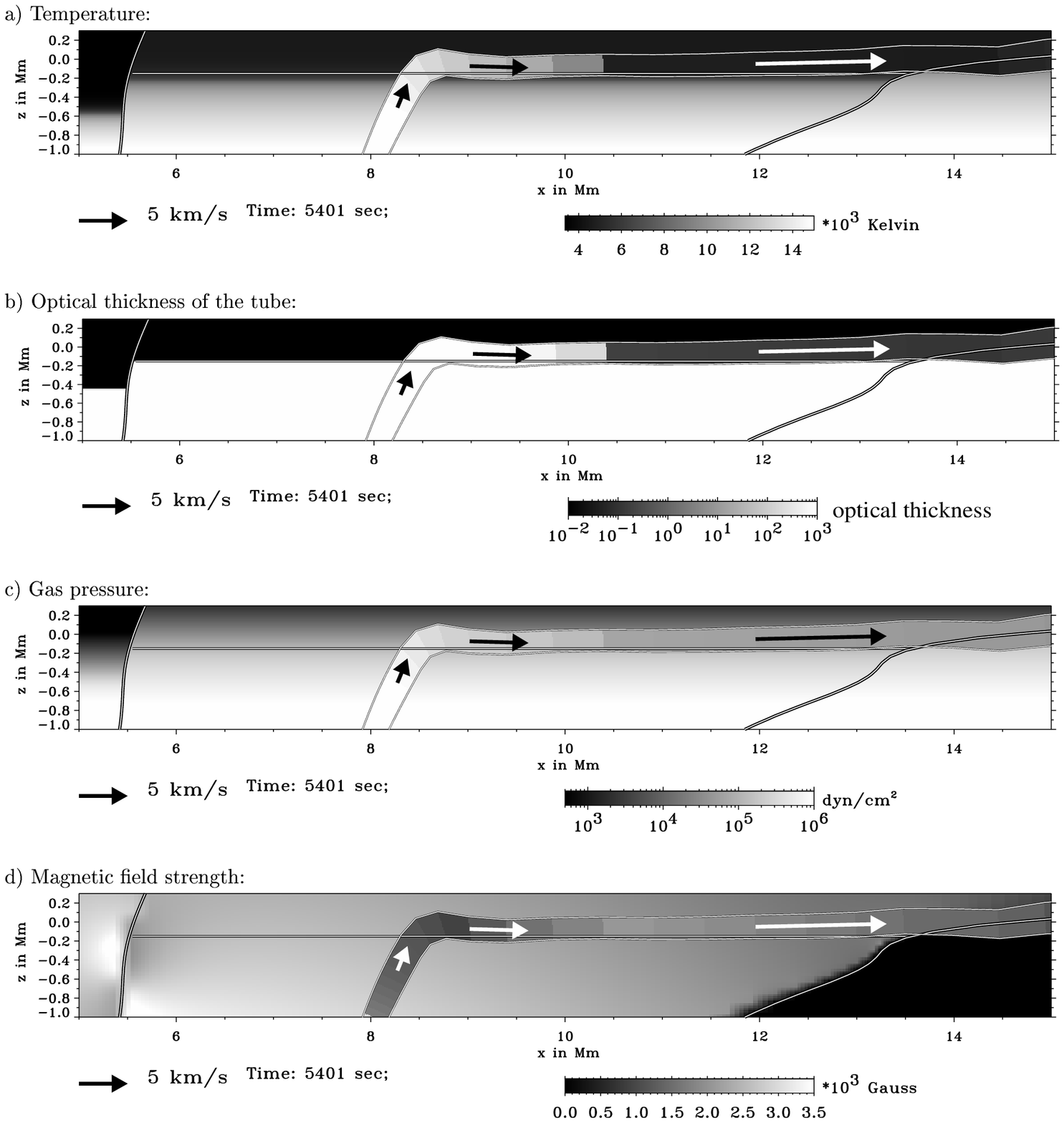}}
 \vspace{0.5cm}
 \caption[]{{\bf a}--{\bf d}.\label{snapshot}
 {\bf Snapshot of the evolution at $t=$ 5\,400 s.} From left to right
 one sees the umbra, the penumbra, and the quiet sun. The tube that has
 risen from the magnetopause lies horizontally above the photosphere,
 being elevated some 100 km. The tube's diameter is magnified by a
 factor 6 for better visibility. As indicated by the legends, the gray
 coding represents physical variables along the tube and in the
 background: {\bf a} Temperature, {\bf b} Optical thickness of the tube,
 {\bf c} Gas pressure, {\bf d} Magnetic field strength.  The arrows in
 the tube visualize longitudinal velocities in the tube. In the lower
 left corner of each plot an arrow with a length that corresponds to 5
 km~s$^{-1}$ is shown.}
 \end{figure*}

In order to discuss the observational consequences of our model, we
concentrate on the evolution of the tube in and above the photosphere,
since only here is the tube observable. Fig. \ref{footpoint} shows the
proper inward velocity of the migrating footpoint as a function of the radial
distance $x$ from sunspot center.  It illustrates the migration of the
tube's footpoint which will be compared with the observed inward
migration of bright penumbral grains. Further, we use
Fig. \ref{snapshot} to discuss the features of our simulations in the
context of the observed penumbral fine structure such as bright and dark
filaments and the Evershed effect. It shows an intermediate stage of
evolution of the tube near the surface layers. From left to right, one
can see the umbra, the penumbra, and the quiet sun being separated by
the peripatopause and the magnetopause, respectively. In
Fig. \ref{snapshot}a, \ref{snapshot}c, and \ref{snapshot}d the gray
coding represents the temperature variation, the gas pressure variation,
and the variation of the magnetic field strength, respectively. Note,
that in each background stratification the temperature and the pressure
vary only with depth, whereas the magnetic field is
two-dimensional. Fig. \ref{snapshot}b shows the optical thickness of
the tube.  Here, the subphotospheric background is illustrated as an
optically thick regime and the atmosphere as an optically thin regime.
For better visibility the diameter of the tube in Fig. \ref{snapshot}
has been magnified by a factor 6. The arrows inside the tube indicate
longitudinal flow velocities. Just below the footpoint a velocity of 3
km~s$^{-1}$ is present, beyond the footpoint the velocity increases to 6
km~s$^{-1}$, and at the outer part of the penumbra near $x=12$ Mm the
flow velocity reaches 13 km~s$^{-1}$.

 \subsection{Penumbral grains and bright filaments}

The footpoint of the flux tube can be identified with a bright penumbral
grain, because the upflow along the tube brings hot plasma to the
surface which appears bright against the darker background.  As
demonstrated in the previous section, the tube's footpoint migrates from
the outer edge of the penumbra inward towards the umbra.  In Fig.\
\ref{footpoint} the data points (plus signs) are equally spaced in time
($\triangle t=300$~s), illustrating that the footpoint starts off at
$x=13\,300$ km with a proper velocity exceeding 2 km~s$^{-1}$. After
1\,200~s (4$^{\rm th}$ data point from the right) the migration of the
footpoint decelerates from 1.5 km~s$^{-1}$ to $\approx 0.1$ km~s$^{-1}$
at $t=7\,200$ s. These values are consistent with the proper motion of
penumbral grains observed by Muller (\cite{Mul73b}).

At the footpoint, the tube's temperature and gas pressure are higher
than the corresponding background values. As the plasma flows along, the
tube loses internal energy by radiation. Thereby it cools and the gas
pressure diminishes. At $x \approx 10$ Mm it reaches temperature
equilibrium with the surroundings. As the gas pressure diminishes, the
magnetic field strength increases according to Eq. (\ref{pressure}), and
flux conservation, Eq. (\ref{flux_conservation}), implies that the
tube's diameter decreases between the footpoint and the point of thermal
equilibrium. Thus, our model is concurrent with observations from Muller
(\cite{Mul73a}, b) and Tritschler et al. (\cite{Tri97}) which
show that penumbral grains consist of a bright coma and a somewhat
dimmer thinner tail which is radially elongated and points away from the
umbra.  Our model suggests that bright filaments are the long tails of
penumbral grains.  However, for lower spatial resolution, a few radially
aligned penumbral grains might also be visible and interpreted as one
bright filament. We want to mention, that the length of the bright tail
depends on the amount of magnetic flux of the tube: More magnetic flux
implies a larger diameter and thus a larger optical depth, a longer
radiative relaxation time (see Eq. (\ref{trad})), and hence a longer
bright tail.

 \subsection{Dark filaments}

In view of Figs. \ref{snapshot}a and \ref{snapshot}b, one can see that
the tube is optically thick as long as it is hotter than the
surroundings. This is due to the temperature dependence of the H$^-$
opacity which is the dominant opacity near the photosphere:
$\tilde\kappa \propto T^{9}$. Another important feature was already
mentioned in Sect. 3.2.2: The tube is elevated above the
photosphere. That means that bright filaments are optically thick
structures overlying a darker background and surrounded by an optically
thin atmosphere.  Therefore, we propose that dark filaments might not
exist {\it per se}. They rather appear as a consequence of spacing
between radially elongated bright filaments (Schlichenmaier et
al. \cite{Sli98}). This is consistent with the statement of Muller
(\cite{Mul73a}, \cite{Mul73b}) that within the penumbra bright features
{\em show up} against a dark background.

 \subsection{Evershed effect}

Our model also offers a consistent explanation for the Evershed effect:
Between the footpoint, at $x\approx 10.5$~Mm, and the outer edge of the
penumbra, the matter inside the tube becomes gradually more transparent.
The optical thickness of the tube decreases to $\tau \approx 10^{-1}$
(see Fig. \ref{snapshot}b) and a line of sight that crosses the tube
reaches optical depth $\tau=2/3$ in the photosphere of the model at
$z=-150$ km, i.e. below the tube.  In other words, after the plasma has
cooled reaching temperature equilibrium with the surroundings, the tube
becomes transparent. The tube constitutes a flow channel that is
elevated approximately 100 km above the photosphere and confines an
outward flow reaching velocities up to $13$ km~s$^{-1}$.  In the
following we discuss the influence of such a flow channel on the line
profiles of photospheric absorption lines.

 \begin{figure}
 \resizebox{\hsize}{!}{\includegraphics{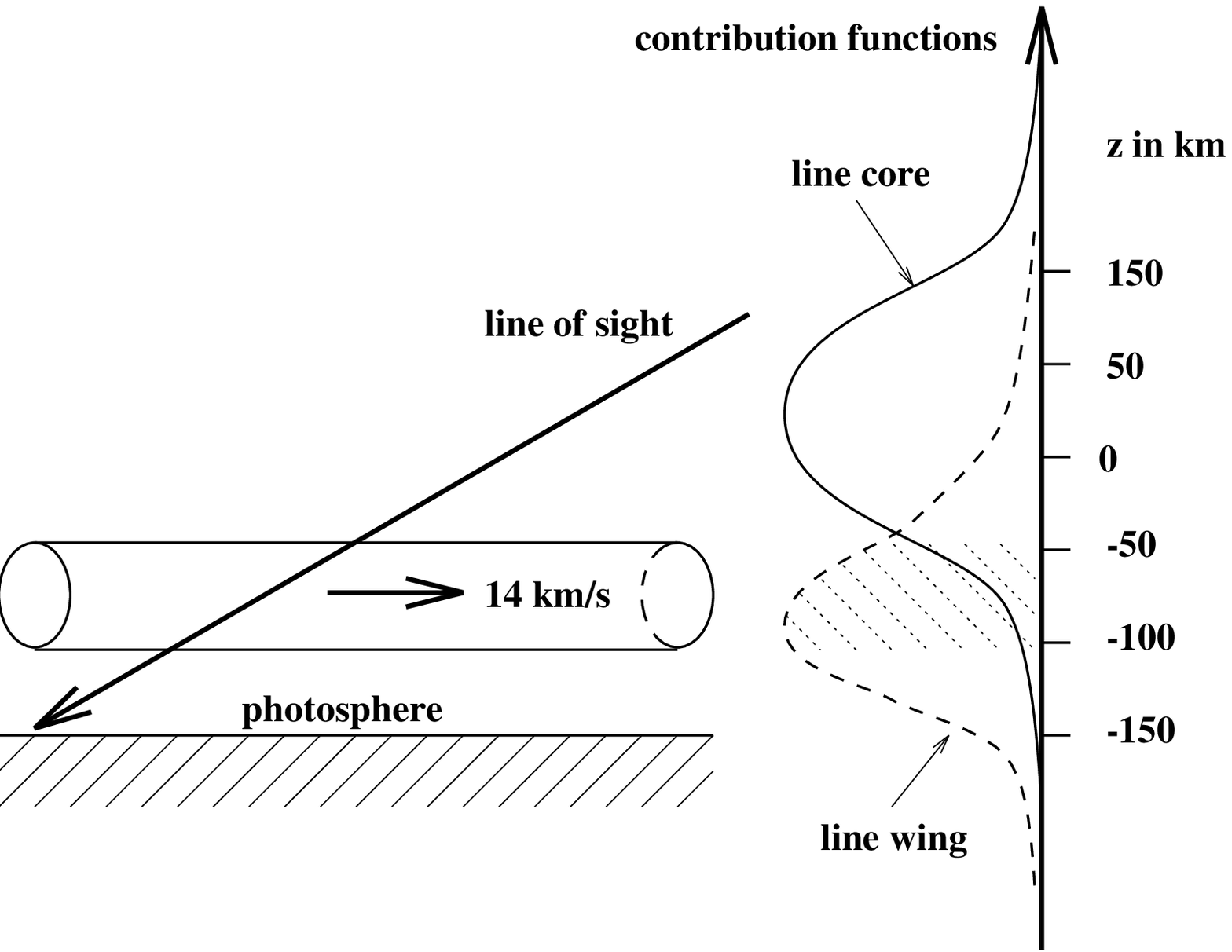}}
 \caption[]{\label{lowflowchannel}
 {\bf Explanation of line asymmetries.} A thin flow channel only affects
 parts of a typical photospheric absorption line. As it is drawn in this
 figure, the flow channel Doppler-shifts a fraction of the contribution
 function for the line wing, but does not influence the contribution
 function of the line core, leaving the line core unshifted. Hence, a thin
 flow channel can naturally explain the observed line asymmetry which is a
 characteristic feature of the Evershed effect.}
 \end{figure}

Generally speaking, a flow channel Doppler-shifts a line profile that
is formed inside the flow channel. In order to understand the line
asymmetry which is a characteristic feature of the Evershed effect, one
should recall that the contribution function of the line core has its
maximum higher up in the atmosphere than the contribution function of
the line wing, as it is depicted for a typical photospheric absorption
line in Fig. \ref{lowflowchannel}. Since the flow channel might be
substantially thinner than the width of the corresponding contribution
function, only distinct parts of a line profile might be shifted.
Therefore, a line asymmetry would be caused naturally by a thin flow
channel. A line asymmetry which is such that the line wing is more
shifted than the line core (see Fig. 1b of Degenhardt \cite{Deg93}) can
be explained by a flow channel that is sufficiently thin and is elevated
only slightly above the photosphere (Schlichenmaier et
al. \cite{Sli98}): Such a flow channel affects the contribution function
of the line wing, but does not influence the contribution function of
the line core which originates in still higher layers.

The diameter of the tube in our simulations is less than 50 km,
depending on the strength of the magnetic field. This has to be taken
into account, when shifted asymmetric line profiles are interpreted in
order to deduce a flow velocity. Wiehr (\cite{Wie95}) deduces a flow
velocity of $\ge$ 5 km~s$^{-1}$ by assuming that the line asymmetry is
caused by a spatially unresolved line satellite being caused by a flow
channel. If one further takes into account, that a typical contribution
function has a full width at half maximum of more than 100 km and that
our simulated tube has a diameter of less than $50$ km, a local flow
velocity exceeding 10 km~s$^{-1}$ may still be consistent with
observation.

 \subsection{Outward flow in bright filaments}

In the preceding section we have demonstrated that the outward flow in
the transparent cooler part of the tube can reproduce a line profile
that is shifted and asymmetric.  However, our model also predicts an
outward flow in optically thick bright filaments. This result is
somewhat surprising, since it is observed that the Evershed flow is
correlated with dark filaments (e.g., Beckers \& Schr\"oter
\cite{BuS69}; Title et al. \cite{Tit93}; Wiehr \& Degenhardt
\cite{WuD94}) and that bright filaments are anti-correlated with the
Evershed flow (Shine et al. \cite{Shi94}).
We explain this apparent contradiction by the fact that all these
measurements were done using Fe\,I lines that have a low excitation
potential of less than 4 eV.
Since these lines originate almost exclusively at atmospheric
temperatures of, say, less than 6000 K, we expect that a hot, optically
thick flow channel has a negligible contribution to the intensity
profile of these lines. If anything, a hot flow channel would at most shift
the line wing, where the line intensity is close to the continuum
intensity. Hence, a hot flow channel may not be detectable with
absorption lines of low excitation potential.
On the other hand, absorption lines which are sensitive to high
temperatures, i.e. lines which have a higher excitation potential, are
expected to originate mainly within a hot flow channel. Hence, observing
bright filaments, i.e. hot plasma, it is more appropriate to use a
``hot'' absorption line with a high excitation potential, such as the
C\,I\,538.03\,nm line. Since this line has an excitation potential of
7.68 eV, its mere existence within the penumbra indicates that hot
plasma is present. In fact, high resolution measurements in the
C\,I\,538.03\,nm line demonstrate the existence of this line within the
penumbra and furthermore, they show line core shifts (Rimmele
\cite{Rim95}, Stanchfield et al. \cite{Sta97}).  Fig. 5 in Stanchfield
et al. shows line core shifts corresponding to Doppler velocities up to
2\,km\,s$^{-1}$ within the penumbra, being blue-shifted on the center
side and red-shifted on the limb side. We take this observation as an
indication for the existence of outward flows in bright filaments.

Finally, we want to mention that Wiehr (\cite{Wie95}) and Balthasar et
al. (\cite{Bal97}) find indications for the existence of a non-zero
velocity in bright filaments, if line profiles are interpreted in terms
of a largely shifted line satellite that stems from dark filaments and a
weakly shifted main component that stems from bright filaments.

In order to decide whether a line shift would be produced by the hot
part of our model tube, one has to compute synthetic line profiles.
Therefore the temperature gradient across the tube is needed, which
forms when an optically thick, hot tube radiates in a cooler, optically
thin atmosphere. Thus, for computing synthetic line profiles, it is
essential to treat the radiative transport for our model tube.

In summary, our model suggests that there are three different structures
within a penumbra: a) a bright component, in which the optically thick and
hot flow may only be detectable using ``hot'' absorption lines, b) a
dark component, in which a flow channel of small optical thickness lies
above the dark background, and c) a dark component consisting of the
dark background with no flux tube above it. In the last component one
would see a dark filament with no Evershed flow present, whereas in case
b) one would look at a dark filament which contains the flow. Hence, the
Evershed signal is correlated with some but not all dark filaments.


 \section{Discussion}

We have presented a numerical simulation of the time dependent dynamic
evolution of a thin magnetic flux tube inside a sunspot penumbra. The
results yield an extension of the studies reported by Jahn et
al. (\cite{Jah96}), and Schlichenmaier et al. (\cite{Sli97b}).  Here, we
aim at a more detailed understanding of the physical structure of the
penumbra. Since our model seems to reproduce the observed small-scale
structure of the penumbra, we take this as an indication that it
supports the concept of interchange convection, on which our results are
based. The model suggests:

 \begin{enumerate}

 \item Penumbral grains are the footpoints of inwards migrating flux
 tubes, in which the enhanced brightness is caused by a systematic
 upflow. At the footpoint an upflow of 3 km~s$^{-1}$ is present.

 \item A bright filament, when interpreted as the dimmer and thinner
 tail of a penumbral grain, is the consequence of a hot plasma outflow
 that cools down during horizontal outward motion.  Bright filaments are
 hotter and optically thicker than the surroundings, and are elevated
 relative to the dark background. In bright filaments a plasma flow is
 present having a velocity of typically 6 km~s$^{-1}$.

 \item The dark background is partly eclipsed by bright
 filaments. Hence, the appearance of dark filaments would be the result
 of spacing between bright filaments.

 \item The Evershed flow occurs with velocities up to 13 km~s$^{-1}$ in
 optically thin ($\tau\approx 10^{-1}$) and almost horizontal channels
 which are elevated some 100 km above the photosphere.

 \end{enumerate}

In our model, the magnetic flux tube and the corresponding plasma flow
extends beyond the outer penumbral boundary. At the boundary, the tube
follows the magnetopause, which slowly gains height and reaches $z=700$
km at $x=27$ Mm. There, the boundary condition allows for a steady
outflow of mass.  Thus, our model does not describe magnetic
flux tubes that dive back beneath the photosphere near the outer edge of
the penumbra, as recently reported by Westendorp Plaza et
al. (\cite{Wes97}) and Stanchfield et al. (\cite{Sta97}). Modeling
dynamical flux tubes that bend back down near the outer edge of the
penumbra requires different upper and lower boundary conditions.
Instead, our model describes the part of the Evershed flow which
continues up into the canopy, as it was observed by Solanki et
al. (\cite{Sol94}).

So far, we have only considered the rise of a tube that lies along the
magnetopause initially. Clearly, for an understanding of interchange
convection, we need not only study different initial conditions, but
also the dynamics of sinking magnetic flux tubes. If no sinking tubes
were present, magnetic flux would be transported continuously from the
outer to the inner part of the penumbra. In consequence, the magnetic
pressure would increase in the inner part of the penumbra. We surmise
that at some point, the gradient of the magnetic pressure, $\grad B_{\rm
b}^2/8\pi$, becomes large enough to suppress the further rise and to
initiate the sinking of the flux tube. 

We want to stress that the acceleration process that causes the outward
flow is related to the siphon flow mechanism (Meyer \& Schmidt
\cite{MuS68}), in the sense that a gas pressure gradient accelerates the
plasma. However, in our model the pressure gradient is created locally
within the penumbra: The flow is initiated by the rise of a magnetic
flux tube and is driven by the superadiabatic background
stratification. In the horizontal part of the tube, radiative cooling
sustains the gas pressure gradient that accelerates the plasma from 3
km~s$^{-1}$ near the footpoint up to 14 km~s$^{-1}$ near the outer edge
of the penumbra.  In result, our model does not require strong magnetic
fields at the outer footpoint as in the classical siphon flow
scenario. We find an acceleration mechanism that is caused naturally in
a certain phase of the evolution of an emerging magnetic flux tube.

Since our simulations yield upflows along subphotospheric flux tubes, it
is tempting to speculate that the surplus brightness of the penumbra as
compared to the umbra is caused by these upflows which constitute a
convective heat transport. Thus, one would suppose that penumbral grains
and bright filaments which are fed by these upflows cause small
fractions of the penumbra to be brighter than the umbra.


 \begin{acknowledgements} RS wishes to thank Prof. Haerendel for fruitful
 discussions and for supporting this work. KJ acknowledges the support
 of the KBN by the grant No. 2 P03D 010 12.  \end{acknowledgements}



\begin{thebibliography}{}
   \bibitem[1997]{Bal97} Balthasar H., Schmidt W., Wiehr E., 1997, Sol.
              Phys. 171, 331
   \bibitem[1960]{Bum60} Bumba V., 1960, Izv. Krymk. Astrofiz. Observ. 23, 253
   \bibitem[1969]{BuS69} Beckers J.M., Schr\"oter E.H., 1969, Sol. Phys. 10, 384
   \bibitem[1995]{Cal95} Caligari P., Moreno Insertis F., Sch\"ussler
           M., 1995, ApJ 441, 886
   \bibitem[1976]{Def76} Defouw R.J., 1976, ApJ 209, 266
   \bibitem[1991]{Deg91} Degenhardt D., 1991, A\&A 248, 637
   \bibitem[1993]{Deg93} Degenhardt D., 1993, A\&A 277, 235
   \bibitem[1909]{Eve09} Evershed J., 1909, MNRAS  69, 454
   \bibitem[1988]{Fer88} Ferriz Mas A., 1988, Phys. Fluids 31, 2583
   \bibitem[1989]{FuS89} Ferriz Mas A., Sch\"ussler M., 1989, 
              Geophys.\ Astrophys.\ Fluid Dynamics 48, 217
   \bibitem[1981]{Gro81} Grossmann--Doerth U., Schmidt W., 1981, A\&A 95, 366
   \bibitem[1963]{Hol63} Holmes J., 1963, MNRAS 126, 155 
   \bibitem[1977]{Hue77} Huebner W.F., Merts A.L.,  Magee N.H.Jr., et al.,
              1977, Los Alamos Sci. Lab. Rep. No. LA-6769-M
   \bibitem[1989]{Jah89} Jahn K., 1989, A\&A 222, 264
   \bibitem[1992]{Jah92} Jahn K., 1992, in:  Sunspots, Theory and Observations, 
              eds. J.H. Thomas, N.O. Weiss, Dortrecht, Kluwer, p. 139
   \bibitem[1994]{JuS94} Jahn K., Schmidt H.U., 1994, A\&A 290, 295
   \bibitem[1996]{Jah96} Jahn K., Schlichenmaier R., Schmidt
           H.U., 1996, Astro. Lett. and Communication 34, 59
   \bibitem[1993]{Lit93} Lites B.W., Elmore D.F., Seagraves P., et al.,
           1993, ApJ 418, 928
   \bibitem[1968]{MuS68} Meyer F., Schmidt H.U., 1968,
           Z. f. angew. Math. Mech., 48, 218
   \bibitem[1986]{Mor86} Moreno Insertis F., 1986, A\&A 166, 291
   \bibitem[1973a]{Mul73a} Muller R., 1973a, Sol. Phys. 29, 55
   \bibitem[1973b]{Mul73b} Muller R., 1973b, Sol. Phys. 32, 409
   \bibitem[1992]{Mul92} Muller R., 1992, in: Sunspots, Theory and Observations, 
              eds. J.H. Thomas, N.O. Weiss, Dortrecht, Kluwer, p. 175
   \bibitem[1982]{Pri82} Priest E.R., 1982, Solar Magnetohydrodynamics, 
              D. Reidel Publishing Company, Dortrecht
   \bibitem[1995]{Rim95} Rimmele T.R., 1995, A\&A 298, 260
   \bibitem[1995]{Ruc95} Rucklidge A.M., Schmidt H.U., Weiss N.O., 
           1995, MNRAS 273, 491
   \bibitem[1997]{Sli97a} Schlichenmaier R., 1997, Dissertation, 
              Ludwig-Maximilians-Univ. M\"unchen, Utz-Verlag,
              M\"un\-chen
   \bibitem[1997]{Sli97b} Schlichenmaier R., Jahn K., Schmidt
              H.U., 1997, in: Advances in the physics of sunspots,
              eds. B. Schmieder, J.C. del Toro Iniesta, M. V\'asquez,
              A.S.P. Conf. Ser. Vol. 118, p. 140
   \bibitem[1998]{Sli98} Schlichenmaier R., Jahn K., Schmidt H.U., 1998, 
	      ApJ Letters 493, L121
   \bibitem[1987]{Smi87} Schmidt H.U., 1987, in: The role of fine-scale
              magnetic fields on the structure of the solar atmosphere,
              eds. E.H. Schr\"oter, M. V\'asquez, A.A. Wyller, Cambridge
              University Press, Cambridge, p. 219
   \bibitem[1991]{Smi91} Schmidt H.U., 1991, Geophys. Astrophys. Fluid 
           Dyn. 62, 249
   \bibitem[1992]{Smi92} Schmidt W., Hofmann A., Balthasar H., et al.,
           1992, A\&A 264, L27
   \bibitem[1965]{Sro65} Schr\"oter E.H., 1965, Z. f. Astrophysik 62, 228
   \bibitem[1994]{Shi94} Shine R.A., Title A.M., Tarbell T.D., et al.,
                  1994, ApJ 430, 413
   \bibitem[1994]{Sol94} Solanki  S.K., Montavon C.A.P., Livingston W., 1994, 
              A\&A 283, 221
   \bibitem[1982]{Sol82} Soltau D., 1982, A\&A 107, 211
   \bibitem[1957]{Spi57} Spiegel E.A., 1957, ApJ 126, 202
   \bibitem[1981a]{Spr81a} Spruit H.C., 1981a, A\&A 98, 155
   \bibitem[1981b]{Spr81b} Spruit H.C., 1981b, A\&A 102, 129
   \bibitem[1983]{Sta83} Stachnik R.V., Nisenson P., Noyes R.W., 1983,
           ApJ 271, L37
   \bibitem[1997]{Sta97} Stanchfield II D.C.H., Thomas J.H., Lites B.W.,
            1997, ApJ 477, 485
   \bibitem[1980]{SuW80} Stellmacher G., Wiehr E., 1980, A\&A 82, 157
   \bibitem[1989]{Sti89} Stix M., 1989, The Sun, Springer Verlag, Berlin
   \bibitem[1988]{Tho88} Thomas J.H., 1988, ApJ 333, 407
   \bibitem[1993]{TuM93} Thomas J.H., Montesinos B., 1993, ApJ 407, 398
   \bibitem[1993]{Tit93} Title A.M., Frank Z.A., Shine R.A., et al.,
                   1993, ApJ 403, 780
   \bibitem[1997]{Tri97} Tritschler A., Schmidt W., Kn\"olker
              M., 1997, in: Advances in the physics of sunspots,
              eds. B. Schmieder, J.C. del Toro Iniesta, M. V\'asquez,
              A.S.P. Conf. Ser. Vol. 118, p. 170
   \bibitem[1997]{Wes97} Westendorp Plaza C., del Toro Iniesta J.C., 
           Ruiz Cobo B., et al., 1997, Nat 389, 47
   \bibitem[1995]{Wie95} Wiehr E., 1995, A\&A 298, L17
   \bibitem[1994]{WuD94} Wiehr E., Degenhardt D., 1994, A\&A 287, 625

\end{thebibliography}
\end{document}